\documentclass[journal=jceaax,manuscript=article]{achemso}

\usepackage{chemformula} % Formula subscripts using \ch{}
\usepackage{}
\usepackage{bm}
\usepackage[T1]{fontenc} % Use modern font encodings
\usepackage{subcaption}
\usepackage{amssymb}
\usepackage{multirow}
\usepackage{longtable}
\usepackage{rotating}
\usepackage{hhline}
\usepackage{amsmath}
\usepackage{gensymb}
\usepackage{graphicx}
\usepackage{color}
\usepackage{tikz}
\usetikzlibrary{shapes.geometric}

\author{Nicolas~Hayer}
\affiliation{Laboratory of Engineering Thermodynamics, RPTU Kaiserslautern, Erwin-Schrödinger-Str. 44, 67663 Kaiserslautern, Germany}

\author{Thorsten~Wendel}
\affiliation{Laboratory of Engineering Thermodynamics, RPTU Kaiserslautern, Erwin-Schrödinger-Str. 44, 67663 Kaiserslautern, Germany}

\author{Stephan~Mandt}
\affiliation{Department of Computer Science, University of California, Irvine, CA 92617, USA}

\author{Hans~Hasse}
\affiliation{Laboratory of Engineering Thermodynamics, RPTU Kaiserslautern, Erwin-Schrödinger-Str. 44, 67663 Kaiserslautern, Germany}

\author{Fabian~Jirasek\textsuperscript}
\affiliation{Laboratory of Engineering Thermodynamics, RPTU Kaiserslautern, Erwin-Schrödinger-Str. 44, 67663 Kaiserslautern, Germany}
\email{fabian.jirasek@rptu.de}

\title{Advancing Thermodynamic Group-Contribution Methods by Machine Learning: UNIFAC 2.0}

\begin{document}

%%%%%%%%%%%%%%%%%%%%%%%%%%%%%%%%%%%%%%%%%%%%%%%%%%%%%%%%%%%%%%%%%%%%%
%% The "tocentry" environment can be used to create an entry for the
%% graphical table of contents. It is given here as some journals
%% require that it is printed as part of the abstract page. It will
%% be automatically moved as appropriate.
%%%%%%%%%%%%%%%%%%%%%%%%%%%%%%%%%%%%%%%%%%%%%%%%%%%%%%%%%%%%%%%%%%%%%
%\begin{tocentry}
%
%Some journals require a graphical entry for the Table of Contents.
%This should be laid out ``print ready'' so that the sizing of the
%text is correct.
%
%Inside the \texttt{tocentry} environment, the font used is Helvetica
%8\,pt, as required by \emph{Journal of the American Chemical
%Society}.
%
%The surrounding frame is 9\,cm by 3.5\,cm, which is the maximum
%permitted for  \emph{Journal of the American Chemical Society}
%graphical table of content entries. The box will not resize if the
%content is too big: instead it will overflow the edge of the box.
%
%This box and the associated title will always be printed on a
%separate page at the end of the document.
%
%\end{tocentry}

%%%%%%%%%%%%%%%%%%%%%%%%%%%%%%%%%%%%%%%%%%%%%%%%%%%%%%%%%%%%%%%%%%%%%
%% The abstract environment will automatically gobble the contents
%% if an abstract is not used by the target journal.
%%%%%%%%%%%%%%%%%%%%%%%%%%%%%%%%%%%%%%%%%%%%%%%%%%%%%%%%%%%%%%%%%%%%%

\begin{abstract}
Accurate prediction of thermodynamic properties is pivotal in chemical engineering for optimizing process efficiency and sustainability. Physical group-contribution (GC) methods are widely employed for this purpose but suffer from historically grown, incomplete parameterizations, limiting their applicability and accuracy. In this work, we overcome these limitations by combining GC with matrix completion methods (MCM) from machine learning. We use the novel approach to predict a complete set of pair-interaction parameters for the most successful GC method: UNIFAC, the workhorse for predicting activity coefficients in liquid mixtures. The resulting new method, UNIFAC 2.0, is trained and validated on more than 224,000 experimental data points, showcasing significantly enhanced prediction accuracy (e.g., nearly halving the mean squared error) and increased scope by eliminating gaps in the original model's parameter table. Moreover, the generic nature of the approach facilitates updating the method with new data or tailoring it to specific applications.
\end{abstract}

%%%%%%%%%%%%%%%%%%%%%%%%%%%%%%%%%%%%%%%%%%%%%%%%%%%%%%%%%%%%%%%%%%%%%
%% Start the main part of the manuscript here.
%%%%%%%%%%%%%%%%%%%%%%%%%%%%%%%%%%%%%%%%%%%%%%%%%%%%%%%%%%%%%%%%%%%%%
\section{Main}
Understanding the thermodynamic properties of mixtures is indispensable in chemical engineering and various related disciplines. However, the vast combinatorial diversity of mixtures makes it impossible to study each relevant mixture experimentally, necessitating reliable prediction methods. Group-contribution (GC) methods address this challenge by deconstructing components into structural groups, significantly reducing the number of parameters since the number of structural groups is much smaller than those of individual components. These methods rely on modeling pair interactions between these structural groups to describe mixture behavior. The effectiveness of GC methods hinges on selecting suitable groups and accurately determining their interaction parameters, both of which depend crucially on the database used for method development and parameterization.

Among GC methods, UNIFAC stands out as the most sophisticated and widely adopted approach for predicting activity coefficients in liquid mixtures. Since its introduction in 1975~\cite{Fredenslund.1975}, UNIFAC has undergone continuous refinement and improvement~\cite{SkjoldJorgensen.1979, Gmehling.1982, Macedo.1983, Tiegs.1987, Hansen.1991, Wittig.2003}, becoming integral to industrial process simulations. Available in both public~\cite{Wittig.2003} and commercial~\cite{UNIFAC_TUC.2023} formats, UNIFAC supports diverse applications, including variants like UNIFAC LLE~\cite{Magnussen.1981} for predicting liquid-liquid equilibria. All UNIFAC variants rely on the same equations but differ in the number and type of groups considered and their parameterization. The process of finding suitable UNIFAC parameters was, in the past, sequential and based on a stepwise extension whenever data became available. This tedious process makes it very difficult to modify decisions taken at early steps. 

This study addresses the challenges of updating and improving UNIFAC by leveraging modern computational techniques, aiming to enhance prediction accuracy and expand its applicability across a broader range of components and mixtures.

Throughout this work, we reference the latest published version of UNIFAC. It was trained on a broad data basis focusing on vapor-liquid equilibrium data to develop a widely applicable model, not one for some specific purpose~\cite{Wittig.2003}. It is astonishing that, despite the importance of UNIFAC, this version is about 20 years old. The leading developers of UNIFAC have updated the method since then, but they have not disclosed these updates – they are only available for members of the UNIFAC consortium. One might ask why no one else has updated this important method since then. The answer to this question is undoubtedly related to the considerable effort required to do this when the conventional strategy is used. Another issue is the accessibility of suitable data. For simplicity, we will label the reference version of UNIFAC~\cite{Wittig.2003} as UNIFAC 1.0 here.

UNIFAC describes the molar excess Gibbs energy, $g^\text{E}$, of a mixture as a function of temperature, $T$, and composition. From $g^\text{E}$, the activity coefficients of the components $i$, $\gamma_i$, in the mixture are obtained. UNIFAC contains group-specific parameters, namely, a size parameter ($R_k$) and a surface parameter ($Q_k$), as well as binary pair-interaction parameters (there are two for each group combination $a_{mn} \neq a_{nm}$, which we will often refer to simply as $a_{mn}$ for simplicity). UNIFAC 1.0 considers 54 \textit{main groups}, subdivided into 113 \textit{subgroups}~\cite{Wittig.2003}.

Applying UNIFAC 1.0 to a given mixture requires the following: i) all components of the mixture must be decomposable into the 113 subgroups, ii) the parameters $R_k$ and $Q_k$ must be available for each relevant subgroup $k$, and iii) the pair-interaction parameters $a_{mn}$ must be available for each binary combination of the relevant main groups $m$ and $n$ (all subgroups of a given main group share the same interaction parameters). The group parameters $R_k$ and $Q_k$ are available for all 113 groups~\cite{DDB2023}, but interaction parameters $a_{mn}$ are missing for many pairs of groups. Specifically, numbers for the interaction parameters are only available for 44\% of all pairs of groups; Fig.~S.1 in the Supporting Information illustrates this. The missing pair-interaction parameters, in some cases due to the challenging fitting process and in other cases due to the lack of experimental data for direct fitting, severely hampers the applicability of UNIFAC 1.0 (a single missing relevant parameter prevents the application of the model).

In this work, we introduce a new way of determining the interaction parameters of GC methods based on machine learning. The approach is based on the idea that the pair-interaction parameters can be treated as elements of a square matrix and that, after suitable training, a matrix completion method (MCM)~\cite{A.Ramlatchan.2018, Koren.2009, Salakhutdinov.2008} can be used to calculate all entries. As numbers for all entries are found, the problem of missing parameters does not exist anymore. In the MCM, so-called group features are determined for all groups from a fit to experimental data. The entire data set is considered during the fit, and a well-defined learning algorithm (in our case, a Bayesian one) is applied. This method replaces the sequential, intuitively guided procedure previously used to determine pair-interaction parameters. As the number of features to be determined scales linearly with the number of main groups $N_\mathrm{MG}$ ($\mathcal{O}(N_\mathrm{MG})$), it is much lower than the number of interaction parameters ($\mathcal{O}(N_\mathrm{MG}^2)$). Consequently, the parameterization of the MCM is significantly more robust than a direct fit of the interaction parameters to the experimental data.

From the features of any two groups $m$ and $n$ of interest, the entries of the interaction parameter matrix $a_{mn}$ are found by a simple matrix multiplication, resulting in a complete set of interaction parameters, thus facilitating the prediction of the activity coefficients $\gamma_i$ for any binary mixture given its structural group composition at any temperature and concentration. 

In prior work, we have already employed MCMs for directly predicting thermodynamic properties of binary mixtures~\cite{Jirasek.2020, Jirasek.2020b, Damay.2021, Hayer.2022, Gromann.2022} and also pair-interaction parameters~\cite{Jirasek.2022, Jirasek.2023}, but here we present the first application of that concept to the development of GC methods with direct end-to-end training on several hundred thousand experimental data points.

The result is UNIFAC 2.0, a hybrid model consisting of the framework of the physical UNIFAC model, in which an MCM from machine learning is embedded.

\begin{figure}[H]
    \centering
    \includegraphics[width=\textwidth]{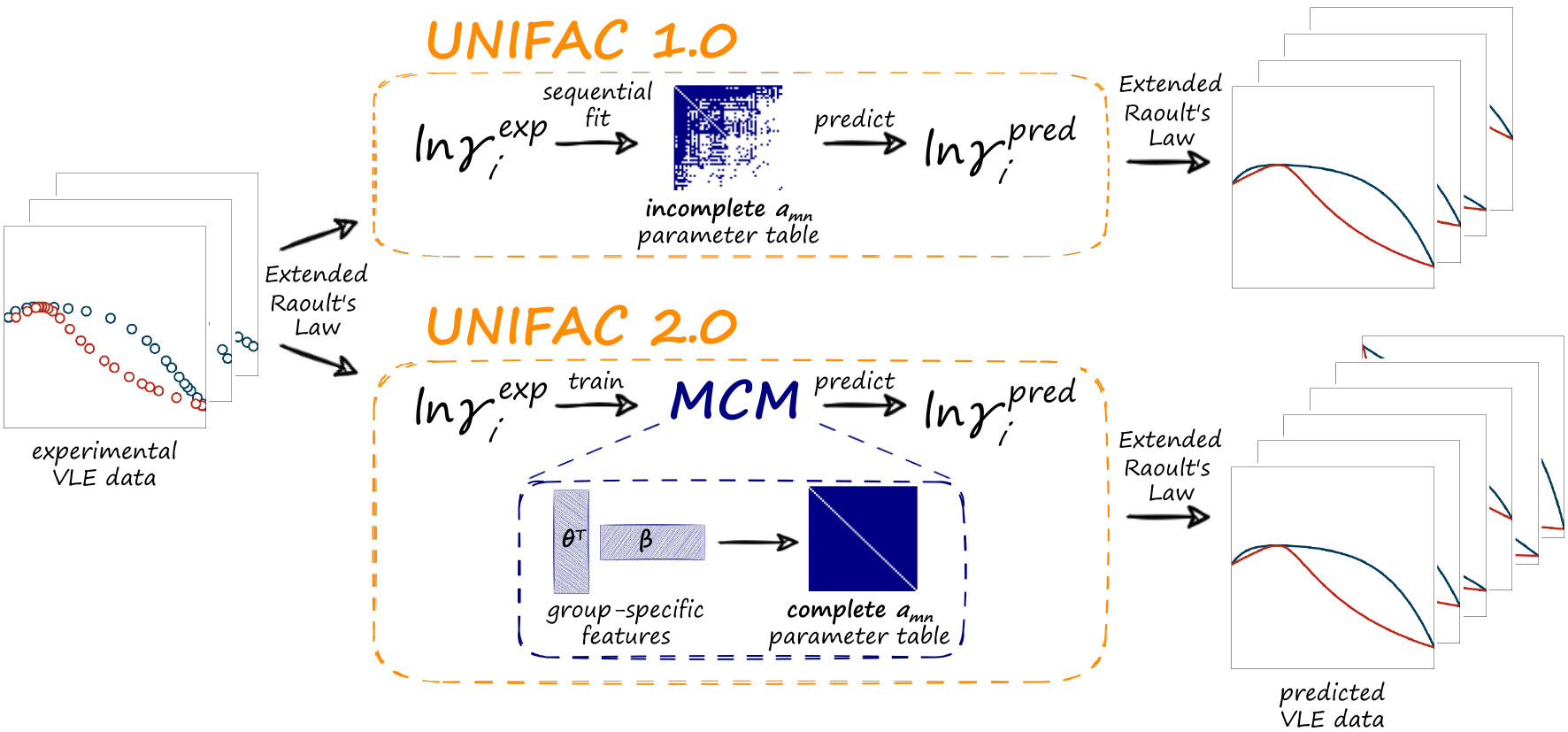}
    \caption{Comparison of UNIFAC 1.0 and UNIFAC 2.0. UNIFAC 1.0 relies on sequential parameter fitting guided by intuition, whereas UNIFAC 2.0 integrates a matrix completion method (MCM) for predicting pair-interaction parameters into the UNIFAC framework. UNIFAC 2.0 is trained end-to-end on experimental logarithmic activity coefficients ($\ln\gamma_i$) derived from binary vapor-liquid equilibrium (VLE) data. After training, the completed pair-interaction parameter matrix facilitates accurate predictions of phase diagrams for a wide range of binary or multi-component mixtures.}
    \label{UNI20_fig:Scheme}
\end{figure}

Fig.~\ref{UNI20_fig:Scheme} compares UNIFAC 1.0 with sequential parameter fitting and UNIFAC 2.0 with end-to-end training of MCM features. Both UNIFAC variants are based on the same structural groups and physical model equations. UNIFAC 2.0 was trained on experimental logarithmic activity coefficients ($\ln\gamma_i$) in binary mixtures derived from vapor-liquid equilibrium data for binary mixtures, cf.~Section "Data" for details.
The MCM is based on the decomposition of the matrix containing the $a_{mn}$ into the product of two feature matrices, thereby enabling the prediction of missing matrix entries through learned features. Each pair-interaction parameter $a_{mn}$ is thereby modeled as follows:
\begin{align}
    a_{mn} = \bm{\theta}_m^\mathrm{T}\cdot\bm{\beta}_n.
    \label{UNI20_eq:matrix_factorization}
\end{align}
Here, $\bm{\theta}_m$ and $\bm{\beta}_n$ are column vectors of length $K$, with $K$ representing the latent dimension, a hyperparameter that was determined in preliminary studies and set to $K=8$.

A Bayesian approach is applied to train the model, treating each logarithmic activity coefficient $\ln\gamma_i$, each feature, and each interaction parameter $a_{mn}$ as a random variable following a probability distribution, detailed further in the Section "Probabilistic Model". From the model training, we obtain a probability density for each $a_{mn}$, the mean of which is used to obtain the scalar value for each parameter. These scalar values are then used in all subsequent evaluations. The completed set of interaction parameters $a_{mn}$, derived from training on all considered binary data, and the subgroup-specific size parameters $R_k$ and $Q_k$ for using UNIFAC 2.0 are provided freely in the Supporting Information. The size parameters are identical to those of the published UNIFAC 1.0 version.

The relevance of the UNIFAC 2.0 becomes apparent when analyzing the applicability of UNIFAC 1.0 and 2.0 considering an example: the Dortmund data bank (DDB), which is the most extensive database for thermodynamic properties, presently lists 39,587 unique components that can be broken down into the published UNIFAC subgroups, which translates into more than 783 million possible binary mixtures. Of these binary mixtures, UNIFAC 1.0 is limited to predicting only 58\% due to missing pair-interaction parameters, whereas UNIFAC 2.0 can be applied to all these mixtures. For multi-component mixtures, the fraction of mixtures that can only be predicted with UNIFAC 2.0 increases dramatically with an increasing number of components, as a mixture drops out if only a single parameter (pair) is missing.

Besides the hybrid model described above, a variant that is based on symmetrical pair-interaction energies $U_{mn} = U_{nm}$ between main groups instead of the asymmetric parameters $a_{mn}$ was developed and tested. The symmetric model has fewer parameters and performs almost as well as the asymmetric model. We report on the asymmetric model here, as it is the standard way to use UNIFAC, and the results can be implemented and used in a very simple manner. Details on the symmetric model are given in the Supporting Information. For a short background discussion of the two variants applied to component-wise pair interactions, see Ref.\cite{Jirasek.2022}.

%\clearpage
\section{Results}
\subsection{Overall Performance of UNIFAC 2.0}
To evaluate the performance of UNIFAC 2.0 and compare it to that of the original UNIFAC 1.0, we employ the mean absolute error (MAE) and the mean squared error (MSE) in the logarithmic activity coefficients $\ln\gamma_i$, which are calculated mixture-wise (from the scores for each binary mixture) to ensure that each mixture is weighted equally in the final score and frequently measured mixtures do not lead to a false impression of the model quality.

In the following, we focus on the predictions of UNIFAC 2.0 obtained after training the hybrid model on all available data points from our database. We have chosen this way for assessing our model since this is likely also the case for UNIFAC 1.0, as the people maintaining UNIFAC and the DDB are essentially the same (although the exact training set of UNIFAC 1.0 has not been disclosed), so we consider the comparison fair. Nevertheless, as described in the following subsections, two additional extrapolation tests were carried out with UNIFAC 2.0 to dispel doubts about its predictive capacity.

\begin{figure}[H]
    \centering
    \includegraphics[width=\textwidth]{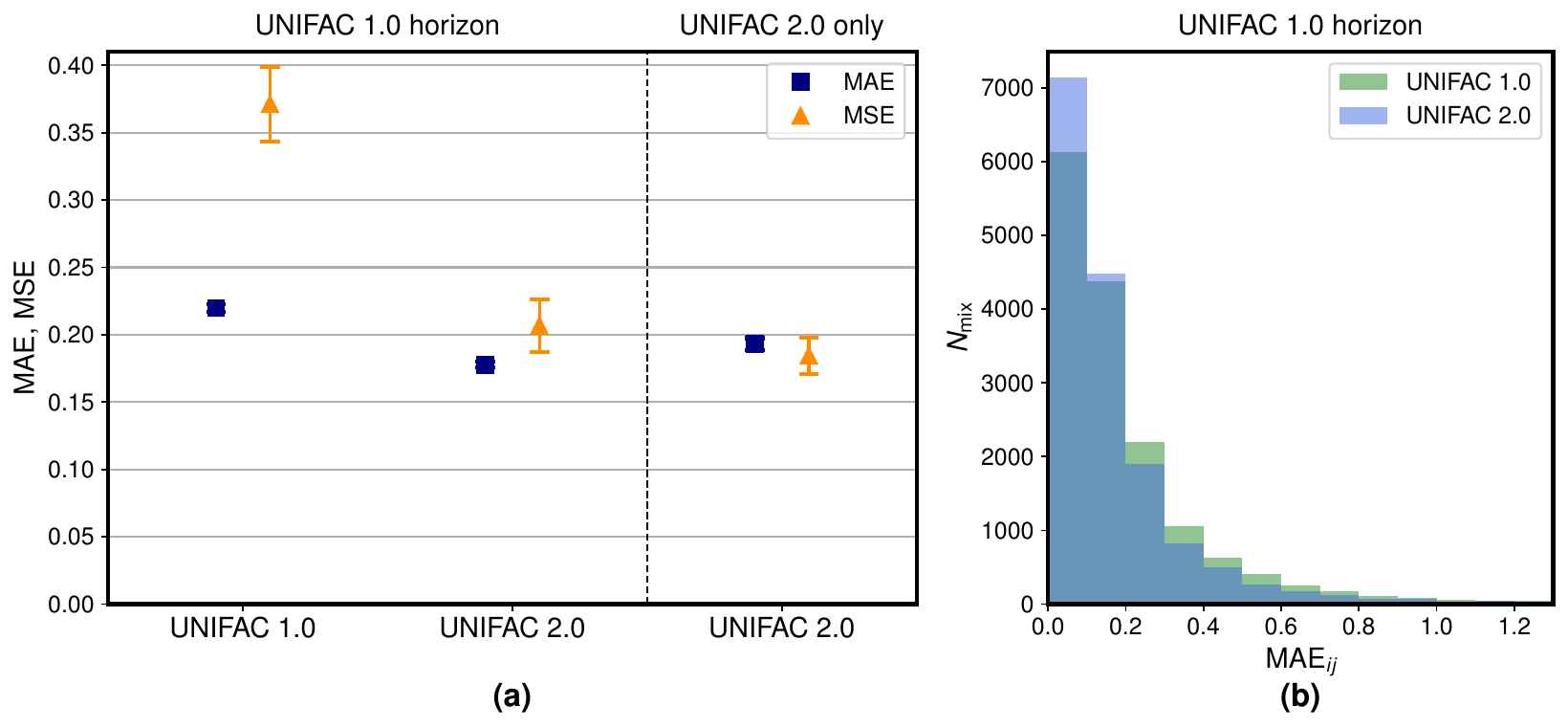}
    \caption{Comparison of results for $\ln\gamma_i$ with UNIFAC 1.0 and UNIFAC 2.0 for different data sets: the "UNIFAC 1.0 horizon" comprises 210,767 data points for 15,758 binary mixtures, while an additional 13,795 experimental data points for 2,957 binary mixtures can only be predicted with UNIFAC 2.0 ("UNIFAC 2.0 only"). (a) Mean absolute error (MAE) and mean squared error (MSE) of the model predictions. Error bars denote standard errors of the means. (b) Histogram of the number of binary mixtures $N_\text{mix}$ that can be predicted with an MAE in a certain interval. The MAE range shown in panel (b) comprises 98.8\% (UNIFAC 1.0) and 99.4\% (UNIFAC 2.0) of all mixtures.}
    \label{UNI20_fig:Performance_bestModel}
\end{figure}

The performance of UNIFAC 2.0 on all available experimental data is shown in Fig.~\ref{UNI20_fig:Performance_bestModel} and compared to UNIFAC 1.0. Since UNIFAC 2.0 has a more extensive scope than UNIFAC 1.0, a distinction is made: all data points that can be predicted with both methods are labeled as the "UNIFAC 1.0 horizon", whereas all data points that can only be predicted with UNIFAC 2.0 are labelled as "UNIFAC 2.0 only".
Fig.~\ref{UNI20_fig:Performance_bestModel}~(a) clearly shows the superior prediction accuracy of UNIFAC 2.0 over UNIFAC 1.0 in both error scores. The MSE can almost be halved compared to the original, demonstrating UNIFAC 2.0's effectiveness in reducing the occurrence of outliers. Even more importantly, the new method not only improves accuracy for data points within the predictive range of UNIFAC 1.0, but it also maintains this accuracy for data points beyond the scope of UNIFAC 1.0, cf.~the results for the "UNIFAC 2.0 only" set.

In Fig.~\ref{UNI20_fig:Performance_bestModel}~(b), a detailed analysis of the MAE for the UNIFAC 1.0 horizon in the form of a histogram of individual binary mixture scores is shown. It underpins that UNIFAC 2.0 achieves an exceptional prediction accuracy: for 7,133 mixtures, the MAE is below 0.1, and thereby in the range of the experimental uncertainty. This accuracy is achieved for only 6,133 mixtures with UNIFAC 1.0.  

\begin{figure}[H]
    \centering
    \includegraphics[width=0.9\textwidth]{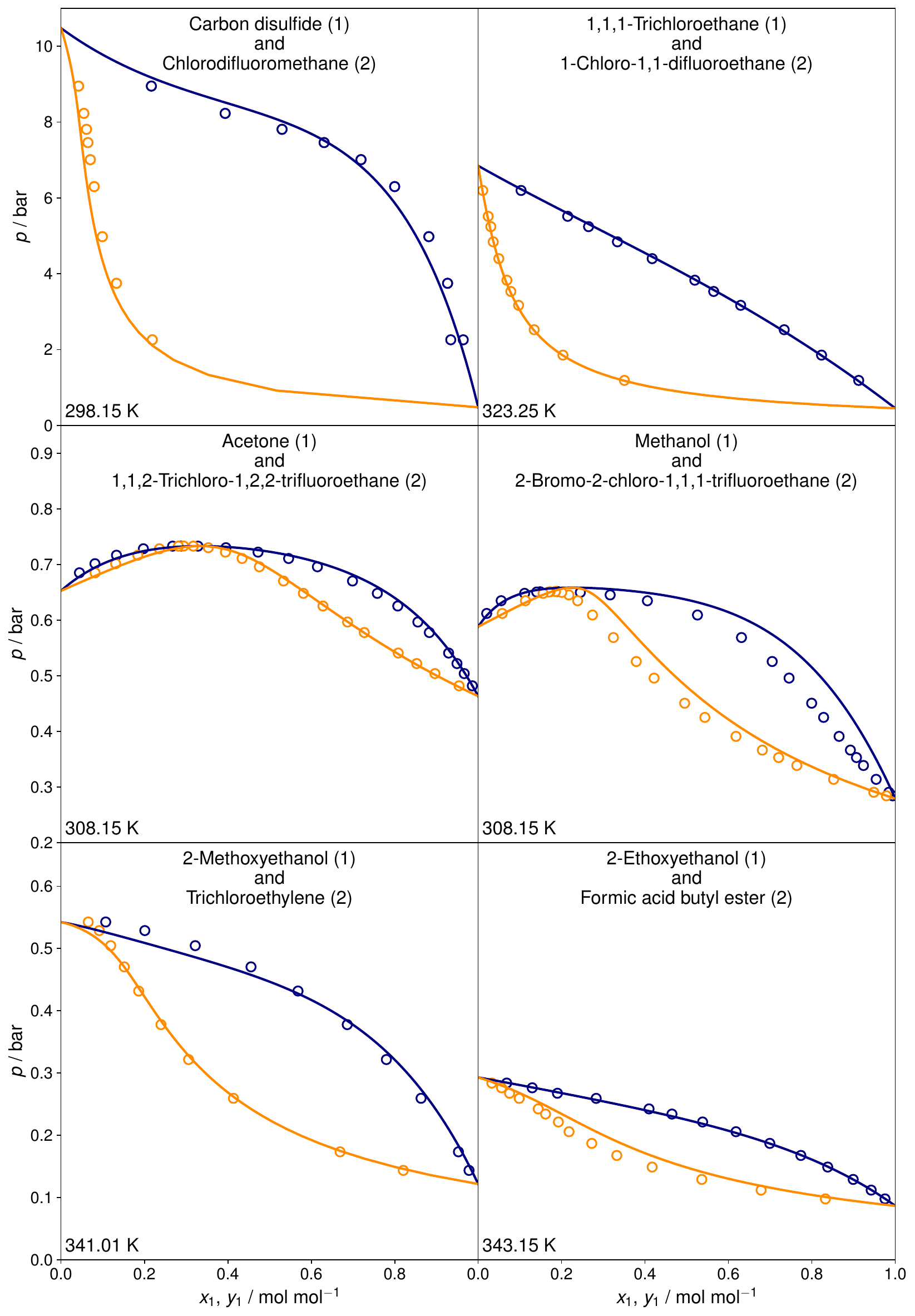}
    \caption{Prediction of isothermal vapor–liquid phase diagrams for binary mixtures with UNIFAC 2.0 (lines) and comparison to experimental data from the DDB (symbols). Blue: bubble point curves. Orange: dew point curves.}
    \label{UNI20_fig:VLE_binary}
\end{figure}

\begin{figure}[H]
    \centering
    \includegraphics[width=\textwidth]{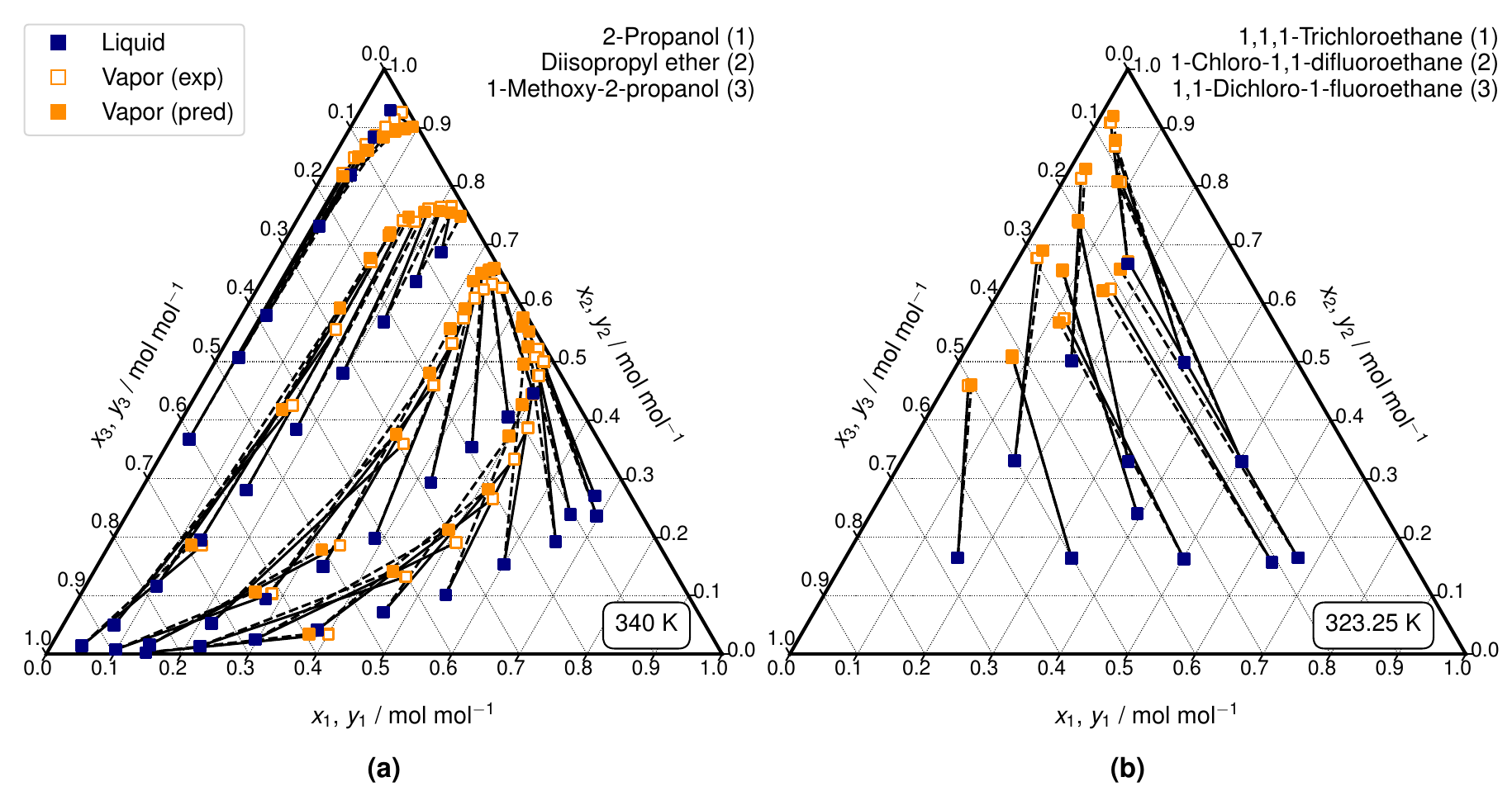}
    \caption{Prediction of isothermal vapor-liquid phase diagrams for ternary mixtures with UNIFAC 2.0 (pred) and comparison to experimental data (exp) from the DDB. The temperature and the composition of the liquid phase were specified, and the composition of the corresponding vapor phase in equilibrium was predicted. Solid lines are experimental conodes, dashed lines are predicted conodes.}
    \label{UNI20_fig:VLE_ternary}
\end{figure}

The activity coefficients obtained by UNIFAC 2.0 can be used directly to predict phase equilibria of mixtures, which are at the core of many tasks in chemical engineering. In Fig.~\ref{UNI20_fig:VLE_binary}, we show six examples of isothermal vapor-liquid phase diagrams predicted by UNIFAC 2.0, cf.~Section "Data" for computational details. All six mixtures are part of the "UNIFAC 2.0 only" set, i.e., they cannot be modeled with the original UNIFAC 1.0. UNIFAC 2.0 accurately describes the phase behavior of all these mixtures. The examples shown in Fig.~\ref{UNI20_fig:VLE_binary} represent typical cases and were selected to cover different types of phase behavior, ranging from small deviations of the ideal behavior to low-boiling azeotropes.
Furthermore, although no data on multi-component mixtures were used for training UNIFAC 2.0, the underlying physical framework of UNIFAC also enables predictions for such mixtures. As examples, Fig.~\ref{UNI20_fig:VLE_ternary} shows isothermal vapor-liquid phase diagrams for two ternary mixtures selected from the "UNIFAC 2.0 only" set, i.e., for UNIFAC 1.0 is not applicable. For each data point, the temperature and the liquid-phase composition (blue symbols in Fig.~\ref{UNI20_fig:VLE_ternary}) were specified and used to predict the corresponding vapor-phase composition in equilibrium with UNIFAC 2.0 (shown as filled orange symbols), which was then compared to the experimentally determined vapor-phase composition (open orange symbols). Excellent accuracy is found.
The results demonstrate the exceptional performance of UNIFAC 2.0, which outperforms UNIFAC 1.0 not only in terms of applicability by closing all gaps in its parameter table but even in terms of prediction accuracy.

% \clearpage
\subsection{Extrapolation to Unknown Components}
\begin{figure}[H]
    \centering
    \includegraphics[width=0.5\textwidth]{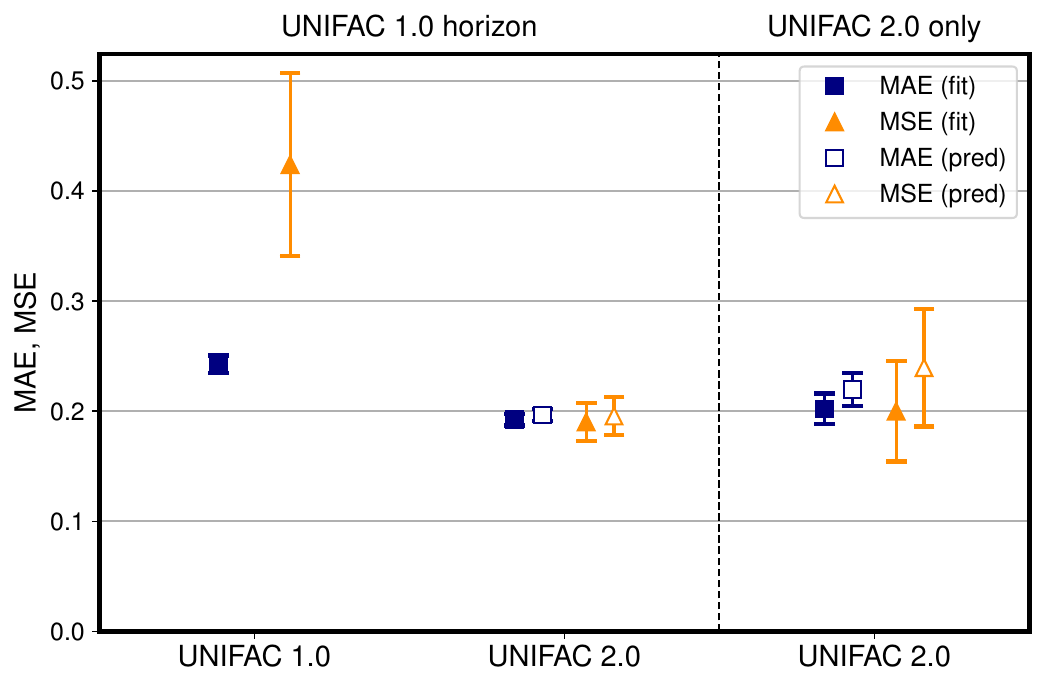}
    \caption{Mean absolute error (MAE) and mean squared error (MSE) of the predicted $\ln\gamma_i$ of mixtures containing unobserved components with UNIFAC 2.0 (pred). For comparison, the results of UNIFAC 2.0 trained on all experimental data and UNIFAC 1.0 are also shown (fit). The "UNIFAC 1.0 horizon" comprises 25,998 data points for 2,202 binary mixtures, while an additional 1,289 experimental data points for 401 binary mixtures can only be predicted by UNIFAC 2.0 ("UNIFAC 2.0 only"). Error bars denote standard errors of the means.}
    \label{UNI20_fig:Performance_unknownComp}
\end{figure}

In a study to evaluate the capacity of UNIFAC 2.0 to extrapolate to unknown components, 100 randomly selected components were intentionally excluded from the training by withholding all data points for systems containing any of these components from the training set and using the systems removed from the training set as the test set. This test set contains 27,287 data points and covers 2,603 different binary mixtures. The results for this test set are shown in Fig.~\ref{UNI20_fig:Performance_unknownComp}, which, again, contains the result from UNIFAC 1.0 for comparison.
Fig.~\ref{UNI20_fig:Performance_unknownComp} shows that the accuracy of the true predictions with UNIFAC 2.0 obtained by withholding the test data during the training (open symbols) is only marginally lower than that of the UNIFAC 2.0 version that was trained on all data points (closed symbols); this holds for both the "UNIFAC 1.0 horizon" and the "UNIFAC 2.0 only" data sets. Furthermore, also in this true predictive test case, UNIFAC 2.0 outperforms UNIFAC 1.0, especially considering the MSE, even though it is likely that UNIFAC 1.0 has been trained on most of the test data points, as discussed above. These findings highlight, on the one hand, the robustness of UNIFAC 2.0 and, on the other hand, the predictive qualities of this hybrid~model.

%\clearpage
\subsection{Extrapolation to Unknown Pair-Interaction Parameters}
\begin{figure}[H]
    \centering
    \includegraphics[width=0.5\textwidth]{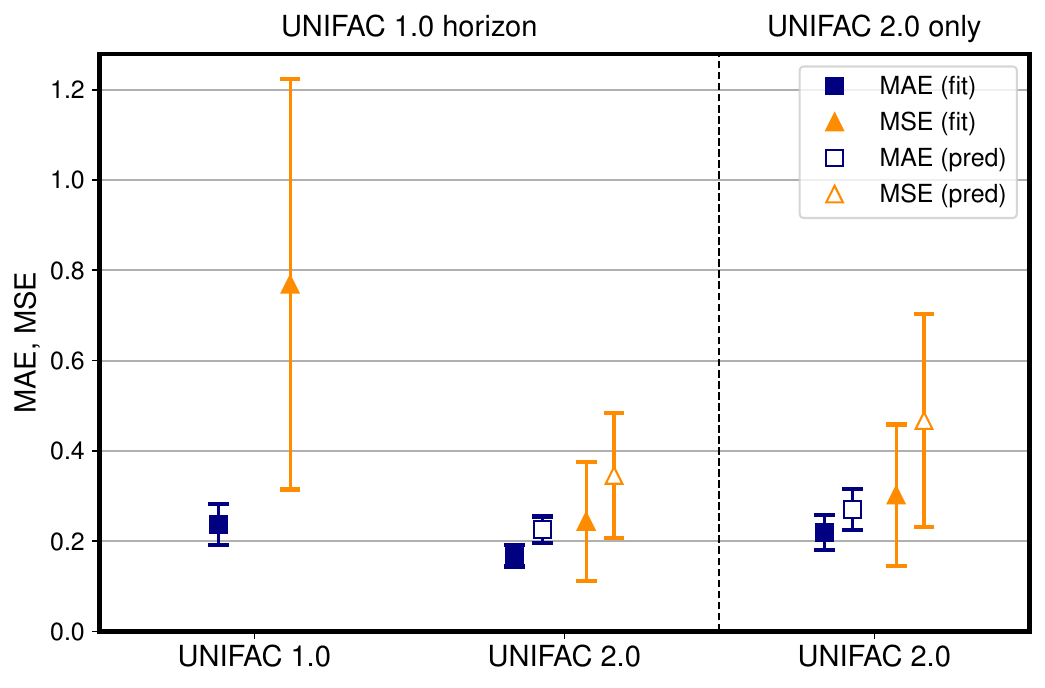}
    \caption{Mean absolute error (MAE) and mean squared error (MSE) of the predicted $\ln\gamma_i$ averaged over 100 test sets with UNIFAC 2.0 (pred). The test sets were created by selecting all data points for which a specific interaction parameter $a_{mn}$ is relevant, cf. Table S.1 in the Supporting Information. The results for UNIFAC 2.0 trained on all experimental data and UNIFAC 1.0 are shown for comparison (fit). Error bars denote standard errors of the means.}
    \label{UNI20_fig:Performance_loo}
\end{figure}

Another, even more challenging, test was carried out by randomly choosing 100 combinations of UNIFAC main groups for which experimental data are available and withholding the data on all systems in which any of the chosen combinations of groups occurs from the training of UNIFAC 2.0. In this way, the capacity of the hybrid model to predict pair-interaction parameters $a_{mn}$ that cannot be obtained by direct fitting is investigated. For each of the 100 selected main group combinations, illustrated in Fig.~S.3 in the Supporting Information, a test set was created that includes the data for those systems in which the selected group combination occurs. All other data points were used to train the model, and the predictions on the test set were evaluated. This process was repeated for all selected main group combinations. MAE and MSE were calculated for each test set. Fig.~\ref{UNI20_fig:Performance_loo} shows the average error scores over all 100 test sets. Again, the results are compared to those of UNIFAC 1.0 and the UNIFAC 2.0 version trained on all data points. Note that the 100 test sets vary strongly in the number of data points and different binary mixtures, as shown in Table~S.1 in the Supporting Information. This table also includes the MAE for each individual test set.
The comparison of the UNIFAC 2.0 predictions to the UNIFAC 1.0 predictions on the "UNIFAC 1.0 horizon" in Fig.~\ref{UNI20_fig:Performance_loo} reveals that the \textit{truly predicted} pair-interaction parameters of UNIFAC 2.0 outperform those of UNIFAC 1.0, which were presumably \textit{fitted} to the experimental data used for evaluation here; this is particularly evident considering the MSE. When comparing the true predictions with UNIFAC 2.0 (open symbols) to those of UNIFAC 2.0 trained on the whole experimental database (full symbols), a slight reduction in prediction accuracy is observed, as expected. However, the differences are small, which demonstrates the robustness of UNIFAC 2.0. The increased standard error associated with the MSE for UNIFAC 1.0 can be attributed to individual test sets for which the predictions are extremely poor.

The results of these tests demonstrate the capability of UNIFAC 2.0 to accurately predict pair-interaction parameters, which enormously increases the scope of this group-contribution method. UNIFAC 2.0 is not only much more applicable than UNIFAC 1.0, but its predictions are also more accurate, as shown by the comparison on the shared horizon. Hence, UNIFAC 2.0 should not only be used when UNIFAC 1.0 cannot be applied, but it should replace UNIFAC 1.0 as the default method for predicting activity coefficients. The fact that UNIFAC 2.0 performs better than UNIFAC 1.0 as measured by lumped criteria, such as the MAE and MSE, that we have used here for describing the performance on a broad database does not exclude, of course, that for specific systems, UNIFAC 1.0 may give better results. Implementing UNIFAC 2.0 is as simple as possible: one must only substitute the original (incomplete) UNIFAC parameter table, e.g., in an established process simulator, with the completed one, which we provide in the Supporting Information. This facility of implementation clearly distinguishes our hybrid model from other machine learning methods for property prediction.

\clearpage
\section{Conclusions}
Group-contribution (GC) methods are widely used workhorses for the prediction of thermodynamic properties of materials. Here, we study how they can be combined with methods from machine learning to obtain hybrid models that outperform their physical parent models. This is demonstrated here for the GC model UNIFAC for predicting activity coefficients in liquid mixtures. UNIFAC is one of the most important GC methods, broadly used in engineering, and implemented in basically all process simulation packages. Like most GC methods for predicting properties of mixtures, UNIFAC is based on the concept of group pair interactions. We demonstrate that these pair interactions can be learned and predicted with matrix completion methods (MCM) from machine learning. The resulting new hybrid model, UNIFAC 2.0, is systematically compared to its physical parent model, UNIFAC 1.0. In contrast to the UNIFAC 1.0 parameter table, which has significant gaps, the parameter table of UNIFAC 2.0 obtained from the MCM has no gaps, leading to a substantial increase in the range of applicability. One could expect to have to pay for this increase in applicability with a deterioration of the accuracy of the predictions - but this is not the case: UNIFAC 2.0 is better than its parent model in both regards. 

The hybrid approach described here also has essential advantages regarding the workflow: as the physical framework is kept, the new model can be implemented very easily in existing software packages; only parameter tables have to be updated to use its advantages. The full UNIFAC 2.0 parameter table is provided in the Supporting Information accompanying this paper. Furthermore, the end-to-end training of the hybrid model to experimental data can be carried out in an automated manner so that updates can be supplied easily if new data become available or targets shift; also, tailored versions of the model, adapted to special needs, can be obtained easily.

\clearpage
\section{Methods}
\subsection{Data}
Experimental data on vapor-liquid equilibria (VLE) and activity coefficients at infinite dilution in binary mixtures were taken from the largest database for thermodynamic properties, the DDB~\cite{DDB2023}. In the preprocessing phase, data points identified as poor quality by the DDB were excluded, and the focus was narrowed to binary mixtures whose components can be decomposed into UNIFAC subgroups. Furthermore, only VLE data points from which the activity coefficients $\gamma_{i}$ of components $i$ in the mixture could be calculated using the extended Raoult's Law
\begin{equation}
    \gamma_{i} = \frac{p\:y_{i}}{p_i^{\text{vap}}\:x_{i}}.
    \label{UNI20_eq:Raoult}
\end{equation}
Here, $p$ is the total pressure and $p_i^{\text{vap}}$ the vapor pressure of component $i$, while $x_{i}$ and $y_{i}$ correspond to the mole fractions of component $i$ in the liquid and vapor phases, respectively.

\subsection{Probabilistic Model}
Our proposed probabilistic model integrates observations ($\ln\gamma_i$) and the latent variables (LVs) that characterize UNIFAC main groups ($\bm{\theta}_m$, $\bm{\beta}_n$) within a Bayesian framework. UNIFAC 2.0 adheres to Bayes' theorem by incorporating three probability distributions: prior, likelihood, and posterior. The prior describes knowledge about the LVs \textit{prior} to fitting the model to the training data. The likelihood constitutes a probability distribution over the observable quantity ($\ln\gamma_i$ here) conditioned on the LVs, i.e., it specifies how the LVs manifest themselves in the data for $\ln\gamma_i$. The aim of Bayesian inference is finding the posterior, which is the probability distribution over the LVs that encapsulates the updated beliefs about the LVs after considering both prior information and empirical data.

Specifically, all $\ln\gamma_i$ and LVs are modeled as independent random variables. A standard normal distribution, i.e., a normal distribution with the mean $\mu=0$ and the standard deviation $\sigma=1$, is used as prior for each LV. The likelihood of observing $\ln\gamma_i$, given the LVs, follows a Cauchy distribution centered around the predicted activity coefficients $\ln\gamma_i^\text{UNIFAC 2.0}$ with scale parameter $\lambda$:
\begin{align}
\label{UNI20_eq:likelihood}
p(\ln\gamma_i\,|\,\bm{\theta}_m,\bm{\beta}_n) &= \text{Cauchy}(\ln\gamma_i^\text{UNIFAC 2.0},\:\lambda),
\end{align}
where $\ln\gamma_i^\text{UNIFAC 2.0}$ is determined via the standard UNIFAC equations~\cite{Wittig.2003} using the predicted interaction parameters $a_{mn}$:
\begin{align}
\label{UNI20_eq:UNIFAC_dependencies}
\ln\gamma_i^\text{UNIFAC 2.0} &= \text{UNIFAC}(a_{mn}, R_k, Q_k, \bm{x}, T).
\end{align}
Here, $R_k$ and $Q_k$ are the subgroup-specific size parameters, $T$ is the temperature, and $\bm{x}$ corresponds to the composition (expressed as mole fractions) of the considered binary mixture.

Written in Pyro, a probabilistic programming language based on Python and PyTorch support~\cite{Bingham.2018}, our probabilistic model adopts stochastic variational inference (VI)~\cite{Blei.2017} for posterior approximation. This approach leverages the Adam optimizer~\cite{Kingma.22.12.2014}, with a learning rate of 0.15. A normal distribution is employed as the variational distribution, with all LVs being treated independently. During training, this approach facilitates learning variational parameters, specifically the mean and standard deviation, for each LV. Based on preliminary studies that have shown robust behavior in terms of predictive performance, the hyperparameters $K = 8$ and $\lambda = 0.4$ were chosen.

Post-training, the LVs inferred from the posterior enable, via Eqs.~(\ref{UNI20_eq:matrix_factorization}) and (\ref{UNI20_eq:UNIFAC_dependencies}), the prediction of $\ln\gamma_i$ for any binary or multi-component mixture, including unstudied ones, whose components can be decomposed in the 113 UNIFAC subgroups.

%%%%%%%%%%%%%%%%%%%%%%%%%%%%%%%%%%%%%%%%%%%%%%%%%%%%%%%%%%%%%%%%%%%%%
%% The "Acknowledgement" section can be given in all manuscript
%% classes.  This should be given within the "acknowledgement"
%% environment, which will make the correct section or running title.
%%%%%%%%%%%%%%%%%%%%%%%%%%%%%%%%%%%%%%%%%%%%%%%%%%%%%%%%%%%%%%%%%%%%%
\clearpage
\begin{acknowledgement}
We gratefully acknowledge financial support by Carl Zeiss Foundation in the project “Process Engineering 4.0”, as well as by Deutsche Forschungsgemeinschaft in the  Priority Program 2363, and in the Emmy Noether Project of FJ. 
\end{acknowledgement}

%%%%%%%%%%%%%%%%%%%%%%%%%%%%%%%%%%%%%%%%%%%%%%%%%%%%%%%%%%%%%%%%%%%%%
%% The same is true for Supporting Information, which should use the
%% suppinfo environment.
%%%%%%%%%%%%%%%%%%%%%%%%%%%%%%%%%%%%%%%%%%%%%%%%%%%%%%%%%%%%%%%%%%%%%

%%%%%%%%%%%%%%%%%%%%%%%%%%%%%%%%%%%%%%%%%%%%%%%%%%%%%%%%%%%%%%%%%%%%%
%% The appropriate \bibliography command should be placed here.
%% Notice that the class file automatically sets \bibliographystyle
%% and also names the section correctly.
%%%%%%%%%%%%%%%%%%%%%%%%%%%%%%%%%%%%%%%%%%%%%%%%%%%%%%%%%%%%%%%%%%%%%
\clearpage
%\bibliography{references.bib}

\providecommand{\latin}[1]{#1}
\makeatletter
\providecommand{\doi}
  {\begingroup\let\do\@makeother\dospecials
  \catcode`\{=1 \catcode`\}=2 \doi@aux}
\providecommand{\doi@aux}[1]{\endgroup\texttt{#1}}
\makeatother
\providecommand*\mcitethebibliography{\thebibliography}
\csname @ifundefined\endcsname{endmcitethebibliography}  {\let\endmcitethebibliography\endthebibliography}{}

\end{document}

%%%%%%%%%%%%%%%%%%%%%%%%%%%%%%%%%%%%%%%%%%%%%%%%%%%%%%%%%%%%%%%%%%%%%
%% The "tocentry" environment can be used to create an entry for the
%% graphical table of contents. It is given here as some journals
%% require that it is printed as part of the abstract page. It will
%% be automatically moved as appropriate.
%%%%%%%%%%%%%%%%%%%%%%%%%%%%%%%%%%%%%%%%%%%%%%%%%%%%%%%%%%%%%%%%%%%%%
%\begin{tocentry}
%
%Some journals require a graphical entry for the Table of Contents.
%This should be laid out ``print ready'' so that the sizing of the
%text is correct.
%
%Inside the \texttt{tocentry} environment, the font used is Helvetica
%8\,pt, as required by \emph{Journal of the American Chemical
%Society}.
%
%The surrounding frame is 9\,cm by 3.5\,cm, which is the maximum
%permitted for  \emph{Journal of the American Chemical Society}
%graphical table of content entries. The box will not resize if the
%content is too big: instead it will overflow the edge of the box.
%
%This box and the associated title will always be printed on a
%separate page at the end of the document.
%
%\end{tocentry}

%%%%%%%%%%%%%%%%%%%%%%%%%%%%%%%%%%%%%%%%%%%%%%%%%%%%%%%%%%%%%%%%%%%%%
%% Start the main part of the manuscript here.
%%%%%%%%%%%%%%%%%%%%%%%%%%%%%%%%%%%%%%%%%%%%%%%%%%%%%%%%%%%%%%%%%%%%%
\section{UNIFAC Parameterization}
\begin{figure}[H]
    \centering
    \includegraphics[width=\textwidth]{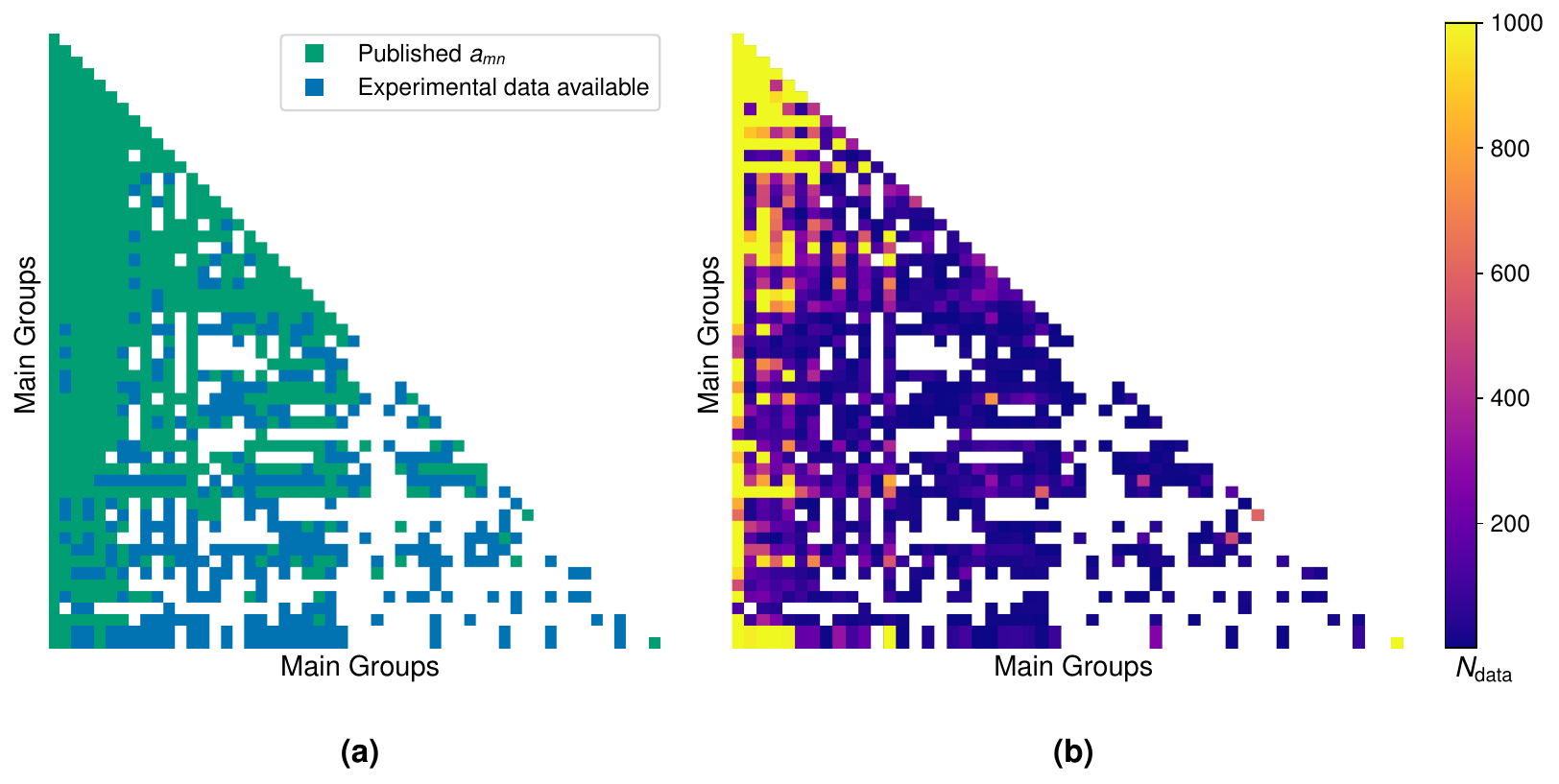}
    \caption{(a) Representation of the published UNIFAC pair-interaction parameters $a_{mn}$~\cite{Wittig.2003} (green) and the ones that could additionally be fitted using the experimental data from the DDB~\cite{DDB2023} (blue). (b) Heatmap of number of experimental data points from the DDB requiring specific $a_{mn}$.}
    \label{UNI20_fig:Heatmaps_DataBase}
\end{figure}

Fig.~\ref{UNI20_fig:Heatmaps_DataBase}~(a) visualizes which pair-interaction parameters $a_{mn}$ are already reported in UNIFAC 1.0 and which $a_{mn}$ can additionally be fitted to the considered database, cf.~Section "Data". The heatmap in Fig.~\ref{UNI20_fig:Heatmaps_DataBase}~(b) indicates the number of experimental data points from the DDB for which a specific $a_{mn}$ is relevant. The figure reveals an extreme heterogeneity, e.g., while 109 $a_{mn}$ (7.6\% of the matrix) is required for at least 1,000 data points, 476 (33\%) are not represented in any available data point.

\begin{figure}[H]
    \centering
    \includegraphics[width=0.5\textwidth]{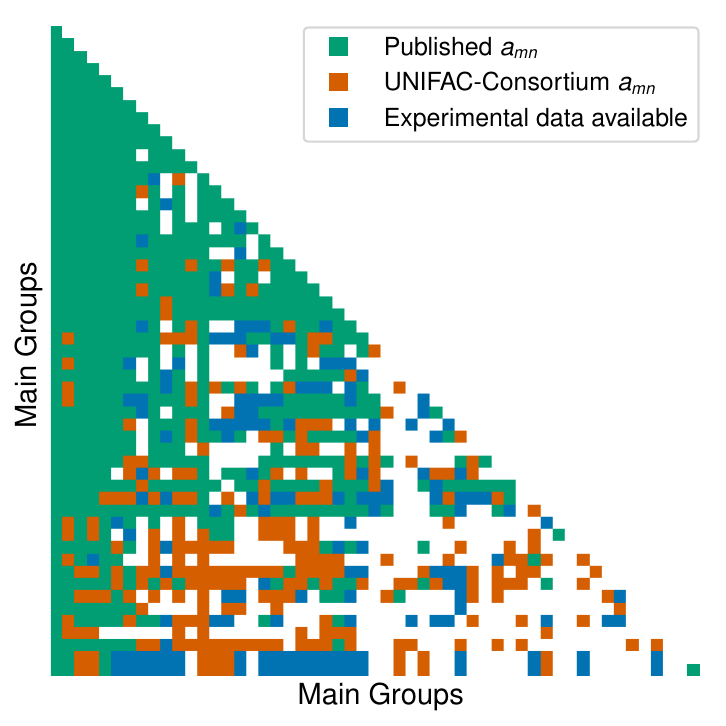}
    \caption{Matrix of existing UNIFAC parameters ($a_{mn}$) of the public UNIFAC 1.0 model~\cite{Wittig.2003} (green), supplemented by those of the commercial UNIFAC-Consortium variant~\cite{UNIFAC_TUC.2023} (orange). Furthermore, group combinations are marked, for which data are available, but no parameters have yet been fitted (blue).}
    \label{UNI20_fig:Heatmap_Consortium}
\end{figure}

Fig.~\ref{UNI20_fig:Heatmap_Consortium} is an extension of Fig.~\ref{UNI20_fig:Heatmaps_DataBase}~(a), additionally including the interaction parameters available for members of the UNIFAC-Consortium. Note that more than the 54 considered main groups are defined for this UNIFAC variant, which are omitted here.
Although the UNIFAC-Consortium model has a substantially increased scope compared to UNIFAC 1.0, Fig.~\ref{UNI20_fig:Heatmap_Consortium} still reveals significant gaps in the interaction parameter matrix, which can be mainly attributed to the lack of available experimental training data. This underlines the importance of methods like UNIFAC 2.0, which are can extrapolate these missing interaction parameters. Since the parameter tables of the UNIFAC-Consortium model are not disclosed, an evaluation and comparison of its predictive accuracy could not be conducted here.

\clearpage
\section{Extrapolation of Unknown Pair-Interaction Parameters}
\begin{figure}[H]
    \centering
    \includegraphics[width=0.5\textwidth]{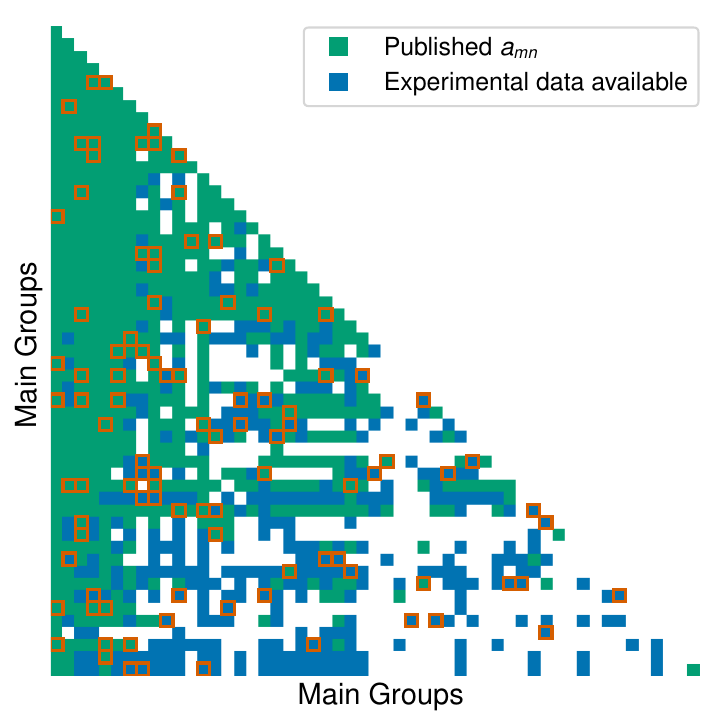}
    \caption{Matrix of existing UNIFAC parameters ($a_{mn}$) of the public UNIFAC 1.0 model~\cite{Wittig.2003} (green) alongside additional group combinations for which experimental data are available (blue). Group-combinations that have been selected for the extrapolation study are highlighted by orange frames, cf.~Table~\ref{UNI20_tab:used_Solutes}.}
    \label{UNI20_fig:Heatmap_loo}
\end{figure}

Fig.~\ref{UNI20_fig:Heatmap_loo} shows the selected group combinations for the extrapolation study. All data points requiring the respective $a_{mn}$ were omitted from the training and used as a test set for each group combination. Since the considered $a_{mn}$ are needed with varying frequencies to predict the binary mixtures of the experimental database, the resulting test sets fluctuate in the number of data points and mixtures. Table~\ref{UNI20_tab:used_Solutes} gives a detailed overview of all 100 test sets.

In this extrapolation study, interaction parameters of UNIFAC 1.0 are available for 62 out of the 100 selected group combinations. However, the availability of these parameters does not guarantee the predictability of all binary mixtures within the test set, as they may need other necessary interaction parameters. To address this distinction, Table~\ref{UNI20_tab:used_Solutes} categorizes the data into two groups: those predictable with both UNIFAC 1.0 and UNIFAC 2.0 ("UNIFAC 1.0 horizon") and those exclusive to UNIFAC 2.0 ("UNIFAC 2.0 only"). Consequently, the prediction of the remaining 38 test sets can solely be carried out with UNIFAC 2.0.
\begin{longtable}{l r r r r r r r}
	\caption{Test sets evaluated for predicting interaction parameters $a_{mn}$. Each set is categorized into two groups: "UNIFAC 1.0 horizon" and "UNIFAC 2.0 only". The structural group identifiers ($m-n$) are identical to UNIFAC 1.0~\cite{Wittig.2003}. The table lists the number of data points ($N_\text{data}$) and binary mixtures ($N_\text{mix}$) for each set. It also includes the mixture-wise mean absolute errors, MAE$^\text{1.0}_\text{mix}$ and MAE$^\text{2.0}_\text{mix}$, for both UNIFAC methods.} \\
	\toprule
    \multirow{2}{*}{$m-n$} & \multicolumn{4}{c}{UNIFAC 1.0 horizon} & \multicolumn{3}{c}{UNIFAC 2.0 only} \\ \cline{2-8}
    & $N_\text{data}$ & $N_\text{mix}$ & MAE$^\text{2.0}_\text{mix}$ & MAE$^\text{1.0}_\text{mix}$ & $N_\text{data}$ & $N_\text{mix}$ & MAE$^\text{2.0}_\text{mix}$ \\
    \midrule
	\endhead
1-17 & 1017 & 134 & 0.32 & 0.22 & 92 & 71 & 0.18\\
1-29 & 392 & 52 & 0.32 & 0.18 & 73 & 25 & 0.32\\
1-32 & 685 & 143 & 0.40 & 0.11 & 89 & 33 & 0.18\\
1-49 & 482 & 32 & 0.36 & 0.16 & 43 & 6 & 0.22\\
1-55 & 1034 & 135 & 1.31 & 0.18 & 202 & 55 & 0.74\\
2-8 & 196 & 39 & 0.24 & 0.33 & 11 & 7 & 0.28\\
2-39 & 259 & 49 & 0.16 & 0.18 &  &  & \\
2-45 &  &  &  &  & 151 & 10 & 0.11\\
3-11 & 2064 & 308 & 0.14 & 0.16 & 110 & 29 & 0.19\\
3-15 & 442 & 43 & 0.26 & 0.27 & 73 & 7 & 0.27\\
3-25 & 4305 & 535 & 0.23 & 0.34 & 292 & 108 & 0.19\\
3-30 & 689 & 22 & 0.17 & 0.11 & 32 & 1 & 0.05\\
3-32 & 135 & 51 & 0.07 & 0.09 & 47 & 24 & 0.18\\
3-39 & 448 & 17 & 0.17 & 0.20 &  &  & \\
3-42 & 65 & 3 & 0.15 & 1.36 & 89 & 4 & 0.19\\
3-43 & 46 & 1 & 0.13 & 0.15 & 86 & 2 & 0.12\\
4-6 & 415 & 19 & 0.21 & 0.15 & 12 & 1 & 0.02\\
4-11 & 1221 & 132 & 0.11 & 0.13 & 76 & 13 & 0.18\\
4-12 & 102 & 7 & 0.17 & 0.12 &  &  & \\
4-48 &  &  &  &  & 75 & 11 & 0.21\\
4-49 & 95 & 6 & 0.05 & 0.07 &  &  & \\
5-6 & 1473 & 39 & 0.28 & 0.18 & 26 & 4 & 0.24\\
5-34 & 112 & 33 & 0.35 & 0.42 & 90 & 20 & 0.26\\
5-49 & 149 & 5 & 0.20 & 0.10 &  &  & \\
5-55 & 35 & 11 & 0.16 & 0.20 &  &  & \\
5-84 & 711 & 100 & 0.19 & 0.13 & 280 & 61 & 0.18\\
6-28 & 37 & 1 & 0.36 & 0.23 &  &  & \\
6-30 & 36 & 1 & 0.19 & 0.14 &  &  & \\
6-32 & 53 & 2 & 0.29 & 0.05 &  &  & \\
7-27 & 94 & 15 & 0.34 & 0.71 & 2 & 2 & 0.09\\
7-39 & 341 & 3 & 0.06 & 0.10 &  &  & \\
7-55 & 15 & 1 & 0.11 & 0.36 &  &  & \\
7-85 &  &  &  &  & 194 & 19 & 0.92\\
8-11 & 99 & 9 & 0.12 & 0.14 &  &  & \\
8-20 & 51 & 6 & 1.41 & 2.57 &  &  & \\
8-28 & 1 & 1 & 0.02 & 0.03 &  &  & \\
8-37 &  &  &  &  & 1 & 1 & 0.13\\
8-38 &  &  &  &  & 30 & 15 & 0.17\\
8-40 &  &  &  &  & 26 & 18 & 0.25\\
8-85 &  &  &  &  & 6 & 1 & 2.77\\
9-10 & 300 & 33 & 0.10 & 0.10 &  &  & \\
9-11 & 995 & 126 & 0.12 & 0.13 & 6 & 2 & 0.04\\
9-20 & 852 & 42 & 0.19 & 0.18 & 9 & 1 & 0.23\\
9-21 & 520 & 54 & 0.30 & 0.30 & 25 & 2 & 0.11\\
9-24 & 251 & 14 & 0.10 & 0.13 &  &  & \\
9-29 & 2 & 2 & 0.10 & 0.08 &  &  & \\
9-38 &  &  &  &  & 38 & 20 & 0.10\\
9-39 & 114 & 10 & 0.09 & 0.14 &  &  & \\
9-40 &  &  &  &  & 131 & 24 & 0.24\\
10-30 &  &  &  &  & 6 & 1 & 1.23\\
10-50 &  &  &  &  & 28 & 15 & 0.10\\
11-12 & 54 & 8 & 0.08 & 0.08 & 3 & 1 & 0.06\\
11-15 & 377 & 1 & 0.09 & 0.03 &  &  & \\
11-30 & 80 & 5 & 0.28 & 0.25 &  &  & \\
11-41 & 394 & 123 & 0.20 & 0.22 & 5 & 3 & 0.33\\
11-48 &  &  &  &  & 45 & 28 & 0.39\\
12-19 & 12 & 1 & 0.09 & 0.03 &  &  & \\
13-26 & 22 & 12 & 0.15 & 0.13 & 27 & 9 & 0.11\\
13-34 & 10 & 9 & 0.08 & 0.07 & 96 & 21 & 0.12\\
13-41 & 587 & 116 & 0.19 & 0.21 & 33 & 11 & 0.59\\
13-85 &  &  &  &  & 1594 & 315 & 0.15\\
14-19 & 23 & 5 & 0.36 & 0.26 &  &  & \\
14-35 & 15 & 1 & 0.26 & 0.28 &  &  & \\
14-41 &  &  &  &  & 24 & 22 & 0.29\\
14-43 & 27 & 3 & 0.14 & 0.15 &  &  & \\
15-24 & 52 & 2 & 0.09 & 0.10 &  &  & \\
15-49 &  &  &  &  & 9 & 1 & 0.07\\
16-32 &  &  &  &  & 1 & 1 & 0.17\\
16-34 &  &  &  &  & 10 & 10 & 0.10\\
18-25 & 9 & 9 & 0.07 & 0.06 &  &  & \\
18-32 &  &  &  &  & 3 & 3 & 0.12\\
18-38 & 32 & 3 & 0.22 & 0.24 &  &  & \\
18-48 &  &  &  &  & 31 & 1 & 0.42\\
19-21 & 150 & 21 & 0.16 & 0.20 & 1 & 1 & 0.24\\
19-35 &  &  &  &  & 40 & 1 & 0.27\\
20-33 & 103 & 2 & 0.14 & 0.13 &  &  & \\
20-34 &  &  &  &  & 12 & 4 & 0.10\\
20-46 & 154 & 3 & 0.39 & 0.37 &  &  & \\
22-55 &  &  &  &  & 5 & 1 & 0.18\\
23-25 & 25 & 3 & 0.06 & 0.07 &  &  & \\
23-30 & 14 & 1 & 0.26 & 0.17 &  &  & \\
23-45 &  &  &  &  & 27 & 4 & 0.09\\
24-45 &  &  &  &  & 2 & 1 & 0.04\\
25-39 & 35 & 2 & 0.45 & 0.45 &  &  & \\
25-46 &  &  &  &  & 2 & 2 & 0.21\\
26-30 &  &  &  &  & 2 & 1 & 0.22\\
27-38 &  &  &  &  & 193 & 104 & 0.11\\
28-37 & 35 & 1 & 0.09 & 0.04 &  &  & \\
30-50 &  &  &  &  & 2 & 1 & 0.05\\
31-32 &  &  &  &  & 1 & 1 & 0.44\\
31-47 & 36 & 1 & 0.10 & 0.21 & 43 & 3 & 0.16\\
32-50 &  &  &  &  & 2 & 2 & 0.64\\
33-38 &  &  &  &  & 43 & 8 & 0.09\\
35-37 &  &  &  &  & 3 & 3 & 0.59\\
38-47 &  &  &  &  & 23 & 15 & 0.13\\
39-47 &  &  &  &  & 9 & 1 & 0.02\\
40-41 &  &  &  &  & 160 & 70 & 0.23\\
41-42 &  &  &  &  & 3 & 2 & 0.22\\
41-51 &  &  &  &  & 9 & 2 & 0.28\\
47-48 &  &  &  &  & 20 & 7 & 0.18\\
\bottomrule
\label{UNI20_tab:used_Solutes}
\end{longtable}

\clearpage
\section{Symmetric UNIFAC Model}
\begin{figure}[H]
    \centering
    \includegraphics[width=0.5\textwidth]{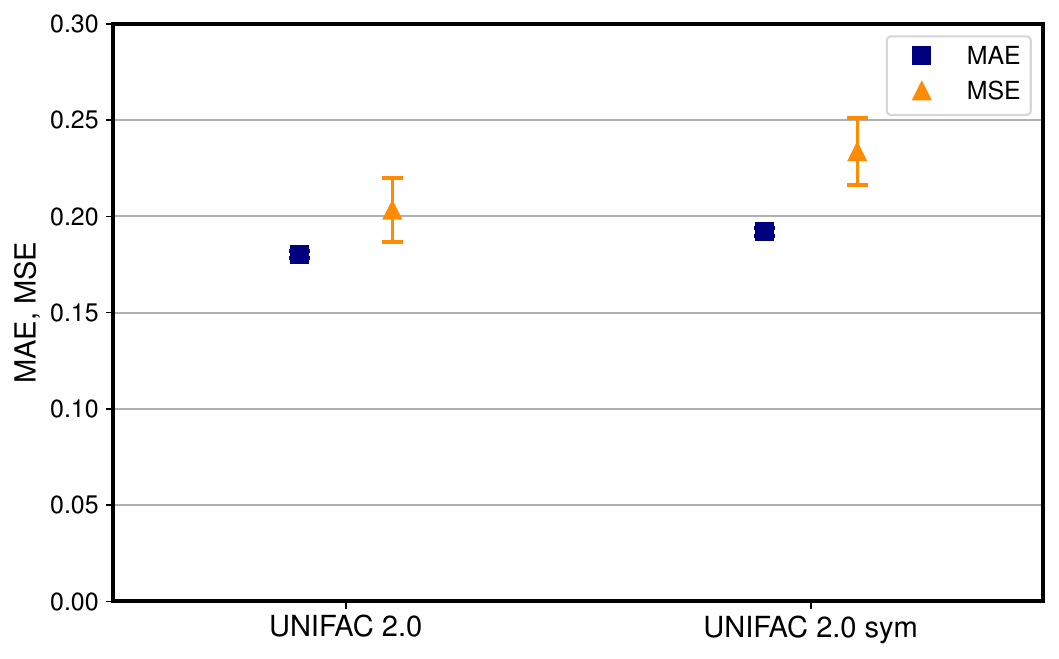}
    \caption{Mean absolute error (MAE) and mean squared error (MSE) of the predicted $\ln\gamma_i$ with UNIFAC 2.0 and a model variant enforcing symmetric interaction energies (UNIFAC 2.0 sym). The whole binary data set was considered, comprising 224,562 data points for 18,715 binary mixtures. Error bars denote standard errors of the means.}
    \label{UNI20_fig:Performance_bestModel_sym_dia}
\end{figure}

In the following, we describe a modification to the UNIFAC method by considering the symmetry of pair-interaction energies, denoted as $U_{mn} = U_{nm}$. This contrasts with the approaches of UNIFAC~1.0 and 2.0, which directly optimize asymmetric pair-interaction parameters ($a_{mn} \neq a_{nm}$) that are derived as follows:
\begin{align}
    a_{mn} &= U_{mn} - U_{nn}, \\
    a_{nm} &= U_{nm} - U_{mm}.
\end{align}
This variant is called \textit{UNIFAC 2.0 sym} and optimizes the symmetric interaction energies, aligning with the physical consistency highlighted in our previous work~\cite{Jirasek.2023}. The predictive performance of this approach is depicted in Figure~\ref{UNI20_fig:Performance_bestModel_sym_dia}, comparing the mean absolute error (MAE) and mean squared error (MSE) for both UNIFAC 2.0 and UNIFAC 2.0 sym across the extensive dataset of 18,715 binary mixtures.
Although the symmetric model offers greater physical consistency, its reduced flexibility slightly impacts predictive accuracy. Therefore, we primarily focus on UNIFAC 2.0 in this work.

%%%%%%%%%%%%%%%%%%%%%%%%%%%%%%%%%%%%%%%%%%%%%%%%%%%%%%%%%%%%%%%%%%%%%
%% The appropriate \bibliography command should be placed here.
%% Notice that the class file automatically sets \bibliographystyle
%% and also names the section correctly.
%%%%%%%%%%%%%%%%%%%%%%%%%%%%%%%%%%%%%%%%%%%%%%%%%%%%%%%%%%%%%%%%%%%%%
\clearpage
% \bibliography{references.bib}

\providecommand{\latin}[1]{#1}
\makeatletter
\providecommand{\doi}
  {\begingroup\let\do\@makeother\dospecials
  \catcode`\{=1 \catcode`\}=2 \doi@aux}
\providecommand{\doi@aux}[1]{\endgroup\texttt{#1}}
\makeatother
\providecommand*\mcitethebibliography{\thebibliography}
\csname @ifundefined\endcsname{endmcitethebibliography}  {\let\endmcitethebibliography\endthebibliography}{}